\documentclass[aps,showpacs]{revtex4}
\begin{document}

\title{Perturbations of an exact solution for 2+1 dimensional critical collapse}

\author{David Garfinkle
\thanks {Email: garfinkl@oakland.edu}}
\affiliation{
\centerline{Department of Physics, Oakland University,
Rochester, Michigan 48309}}

\author{Carsten Gundlach
\thanks {Email: C.Gundlach@maths.soton.ac.uk}}
\affiliation{
\centerline{Faculty of Mathematical Studies, University of Southampton,
SO17 1BJ, UK}}

\null\vspace{-1.75mm}

\begin{abstract}

We find the perturbation spectrum of a family of spherically symmetric
and continuously self-similar (CSS) exact solutions that appear to be
relevant for the critical collapse of scalar field matter in 2+1
spacetime dimensions. The rate of exponential growth of the unstable
perturbation yields the critical exponent.  Our results are compared
to the numerical simulations of Pretorius and Choptuik and are
inconclusive: We find a CSS solution with exactly one unstable mode,
which suggests that it may be the critical solution, but another CSS
solution which has three unstable modes fits the numerically found
critical solution better.

\end{abstract}
\maketitle
\section{Introduction}

Critical gravitational collapse, as first found by
Choptuik \cite{matt}, has been studied in many
systems \cite{carsten}. For the most part, these studies have been
numerical: generally the critical solution has not been found in
closed form.  However, gravitation in 2+1 dimensions is a much simpler
system than in 3+1 dimensions.  Thus one might hope to find a closed
form solution of 2+1 critical collapse.  Pretorius and
Choptuik \cite{mattandfrans} performed numerical simulations of 2+1
critical gravitational collapse with a massless, minimally coupled
scalar field and a cosmological constant.  A numerical treatment 
of this system
has also been done by Husain and Olivier.\cite{husain}   
One of us \cite{dg} found a
closed form continuously self-similar (CSS) solution of the 2+1
Einstein-scalar equations that agrees with the work of
Ref. \cite{mattandfrans}.  Other closed form solutions of 
these equations have also been treated\cite{marco,clement} and the
approach to the singularity in this system has been analyzed.\cite{burko} 

A critical solution, when perturbed, should have exactly one unstable
mode, which grows as $e^{kT}$ for some constant $k$, where $T$ is a
coordinate such that $\partial/\partial T$ is the homothetic vector of
the CSS critical solution.  In the collapse of a one parameter family
of initial data, a quantity $Q$ with dimension ${ ({\rm length})}^s$
obeys a scaling relation $Q \propto {{|p-{p*}|}^{\gamma s}}$ where $p$
is the parameter and $p*$ is its critical value.  The quantities $k$
and $\gamma$ are related by $\gamma =1/k$.  Thus by treating
perturbations of a critical solution, one could find $k$ and compare
to the numerical value of $\gamma$ found in simulations of near
critical collapse.  Such a perturbation treatment was begun
in \cite{dg} but was not completed due to questions about appropriate
boundary conditions to impose on the perturbations.

In this paper, we complete the perturbation treatment begun
in \cite{dg}.  Throughout we use double null coordinates rather than
the Bondi coordinates used in \cite{dg}.  This tends to clarify
the issue of boundary conditions for the perturbations.  Section II
presents the field equations and background solution in double null
coordinates.  The perturbations are treated in section III.
Conclusions are presented in section IV.

\section{Field equations and background solution}

The simulations of Ref. \cite{mattandfrans} used the
Einstein-scalar equation with cosmological constant.  However, in the
approach to the critical solution, the cosmological term becomes
negligible.  Therefore we approximate the Einstein equations as
\begin{equation}
{R_{ab}} = 4 \pi {\nabla _a} \phi {\nabla _b} \phi.
\label{einsteinscalar}
\end{equation}
We consider an axisymmetric 2+1 metric in double null coordinates,
which takes the form
\begin{equation}
d {s^2} = 2 {e^{2 A}} d u d w + {B^2} d {\theta ^2},
\label{2p1metric}
\end{equation}
where $A$ and $B$ are functions of $u$ and $w$.  We choose $u$ at the
origin to be proper time from the singularity, but place no
restrictions on $w$.  For a metric of the form (\ref{2p1metric}), the
Einstein-scalar equations (\ref{einsteinscalar}) become 
\begin{eqnarray}
2 B {{{\partial ^2} \phi} \over {\partial u \partial w}} + {{\partial B}
\over {\partial u}} {{\partial \phi } \over {\partial w}} + 
{{\partial B} \over {\partial w}} {{\partial \phi } \over {\partial u}}
&=& 0,
\label{wave}
\\
{{{\partial ^2} B}\over {\partial u \partial w}} &=& 0,
\label{esa}
\\
2 {{\partial B} \over {\partial w}} {{\partial A} \over {\partial w}}
- {{{\partial ^2}B} \over {\partial {w^2}}}&=&  4 \pi B {{\left ( 
{{\partial \phi } \over {\partial w}}\right ) }^2} ,
\label{esb}
\end{eqnarray}
(plus two additional components of the field equations which are 
redundant).
The quantity $A$ is fixed at the origin by the requirement that $u$ be
proper time there.  

To present the background solution, it is helpful to introduce coordinates
$T$ and $y$ given by
$-u={e^{-T}}$ and
$y={w^n}/{\sqrt {-u}}$
where $n$ is a constant. There is a family of regular CSS solutions
parameterized by a positive integer $n$ (the $q$ of Ref.
\cite{dg}), with
\begin{eqnarray}
{\bar B} &=& {1 \over 2} {e^{-T}} ( 1 - {y^2}) ,
\label{bbar}
\\
{\bar \phi} &=& c \left [ T - 2 \ln \left ( {{1+y} \over 2}\right ) \right ] ,
\label{phibar}
\\
{\bar A} &=&  {{1} \over 2} \ln n + \left ( 1 - {1 \over {2 n}}\right ) 
\left [ -{{T} \over 2} + 2 \ln \left ( {{1+y} \over 2}\right ) \right ] .
\label{abar}
\end{eqnarray}
Here, the constant $c$ is given by
$c = \pm {\sqrt {(2n-1)/8 \pi n}}$
and a bar denotes a background quantity. The solution with $n=4$
appears to be a good fit to the critical solution found by Pretorius
and Choptuik \cite{mattandfrans} in numerical collapse simulations.

The spacetime is CSS with homothetic vector
$\partial/\partial T$. The log-scale coordinate $T$ used here is the
same as that defined in Ref. \cite{dg}. The self-similarity coordinate
$y$ defined here is related to the coordinate $R$ defined in
Ref. \cite{dg} by $y^2=1-2R$. However, this simple relation between
$y$ and $R$ does not hold when the spacetime is perturbed. Note that
in these coordinates the origin is at $y=1$ while the past light cone
of the singularity is at $y=0$.  Though we have introduced the
coordinates $T$ and $y$ for convenience, the criterion for smoothness
is that $A$, $B$ and $\phi$ be smooth functions of $u$ and $w$. The
background solution is smooth provided that $n$ is a positive integer.

To study the global structure of the background spacetimes we make
another coordinate change $y\equiv x^n$. The metric becomes 
\begin{equation}
ds^2=e^{-2T}\left[\left({1+x^n\over
2}\right)^{4\left(1-{1\over 2n}\right)}\left(2ndx-x\,dT\right)dT
+{1\over 4}\left(1-x^{2n}\right)^2d\theta^2\right].
\end{equation}
The maximal extension of the spacetime is provided by this metric with
$-\infty<T<\infty$ and $-1\le x\le 1$. The Ricci scalar is
proportional to $e^{2T}$, and so $T=\infty$ is a curvature
singularity. The regular center is at $x=1$, and the past lightcone
is at $x=0$. Lines of constant $x$ (trajectories of the homothetic
vector field) are timelike for $x>0$ and spacelike for $x<0$. The
rings of constant $x$ and $T$ are closed trapped surfaces for $x<0$,
and so $x=0$ can be interpreted as an apparent horizon. 

Note that the $x < 0$ region of the spacetime contains
trapped surfaces, while a critical solution (since it lies on the
boundary of those spacetimes with and those without trapped surfaces) 
cannot itself contain a trapped surface.  Thus it is only the $x>0$ part
of the spacetime that matches the corresponding part of the numerical 
critical solution.  That part of the critical solution that is outside the
past light cone of the singularity is not given by our background solution.

In 2+1 dimensions, the cosmological constant is also necessary for black hole
formation, and this goes beyond the approximation $\Lambda=0$ here
(and beyond our current understanding). In a related complication, in
the presence of a cosmological constant, a black hole will always
eventually form and capture all the scalar field matter. The threshold
behaviour and mass scaling studied by Pretorius and Choptuik applies
only to prompt black hole formation by a scalar field pulse that has
not yet reached the AdS timelike null-infinity.

\section{Perturbations}

We now consider perturbations of the background solution.  Here we use
a $\delta$ to denote a perturbed quantity. Any perturbation mode is of
the form
\begin{eqnarray}
\delta B &=& {e^{(k-1)T}} b(y) ,
\label{deltab}
\\
\delta \phi &=& {e^{kT}} H(y),
\label{deltaphi} \\
\delta A &=& e^{kT} a(y),
\label{deltaA}
\end{eqnarray}
where $k$ is a constant. The perturbation grows as the singularity is
approached if $k>0$. 

These perturbations must satisfy the linearized versions of equations
(\ref{wave}-\ref{esb})  which 
we now solve in turn. Substituting the ansatz
(\ref{deltab}) into the perturbation of equation (\ref{esa}), we find
\begin{equation}
y {b''} + (2k-1) {b'} = 0.
\label{deltabeqn}
\end{equation}
The general solution is
\begin{equation}
\label{bsolution}
b = c_0 + {c_1} (1 - {y^{2-2k}}),
\label{deltabsoln}
\end{equation}
for arbitrary constants $c_0$ and $c_1$. We may also ask which
infinitesimal gauge transformations $x^\mu\to x^\mu+\xi^\mu$ create
metric perturbations of the form (\ref{deltab}-\ref{deltaA}). It turns
out that they form a 2-parameter family with the same parameters $c_0$
and $c_1$. The perturbation $b$ is therefore pure gauge. In order to
study regularity at the origin, we introduce Cartesian coordinates
$t=(u-{w^{2n}})/2$ and $\bar r=(-u-{w^{2n}})/2$. The background metric then
approaches the usual Minkowski form at the origin. All metric
coefficients, and the scalar field $\phi$, are regular at $\bar r=0$,
and are even functions of $\bar r$. The gauge transformation $\xi^\mu$
takes the form
\begin{eqnarray}
{\xi ^t} &=& - [(c_0+c_1){{(-t+{\bar r})}^{1-k}} + c_1 {{(-t-{\bar
r})}^{1-k}}],\\ {\xi ^{\bar r}} &=& (c_0+c_1){{(-t+{\bar r})}^{1-k}} -
c_1{{(-t-{\bar r})}^{1-k}}.  
\end{eqnarray}
This is a regular vector field at $r=0$ if and only if ${c_0}=0$, and we
assume this from now on. We are left with a family of linear
perturbation gauges parameterized by $c_1$. We shall later fix $c_1$
by further regularity considerations.

The perturbation of equation
(\ref{wave}) now becomes
\begin{equation}
{1 \over 2} y (1 - {y^2}) {H''} + [ k - (k+1) {y^2} ] {H'} - k y H
= {{2 c {c_1}}\over {{(1+y)}^2}} \left [ k(1+y) - y + {y^{1-2k}} 
(1 - k(1+y)) \right ]  .
\label{deltaphieqn}
\end{equation}
The solution of this equation that is regular at the origin ($y=1$) is 
\begin{equation}
H = -{{ 2 c {c_1}}\over {1+y}} \left ( 1 + {y^{1-2k}} \right ) + 
{c_2} F(k,1/2,1,1-{y^2}), 
\label{deltaphisoln}
\end{equation}
where $c_2$ is a constant and $F$ is a hypergeometric function.  When
${c_2}=0$, the perturbation simply results from applying an
infinitesimal coordinate transformation to the background and is
therefore pure gauge. 

We now address the issue of smoothness of the perturbation at $y=0$,
the past light cone of the singularity.  We first consider the
particular cases $k=1$ and $k=1/2$. In the case $k=1$, we have
$F(1,1/2,1,1-{y^2})=1/y$. Therefore regularity at $y=0$ requires that
$H=0$ for the case $k=1$. In the case $k=1/2$, we have
$F(1/2,1/2,1,1-{y^2}) = (2/\pi) K({\sqrt {1-{y^2}}})$ where $K$ is an
elliptic integral.  The quantity $K({\sqrt {1-{y^2}}})$ diverges at
$y=0$. Since the first term on the right hand side of equation
(\ref{deltaphisoln}) is regular at $y=0$ for $k=1/2$, it follows that
for $k=1/2$ a regular perturbation must have ${c_2}=0$, which is pure
gauge.

Now we consider general values of $k$. Using an identity for
hypergeometric functions (section 9.13 of  \cite{integrals}) we have
\begin{equation}
F(k,1/2,1,1-{y^2})= {{\Gamma ( {\textstyle {1 \over 2}}- k)} \over
{{\sqrt \pi} \, \Gamma (1-k)}} F(k,{\textstyle {1 \over
2}},{\textstyle {1 \over 2}}+k,{y^2}) + {{\Gamma ( k-{\textstyle {1
\over 2}})} \over {{\sqrt \pi} \, \Gamma (k)}} {y^{1-2k}}
F(1-k,{\textstyle {1 \over 2}}, {\textstyle {3 \over 2}} -k,{y^2}).
\label{hyperidentity}
\end{equation}
Here $\Gamma (x)$ denotes the gamma function.  This formula applies
except in the case where $k$ is half of an odd integer. In that case,
one of the two $\Gamma$-functions has a pole. The correct formula then
involves a $\ln y$ term, and is therefore not regular at $y=0$. All
odd half-integer values of $k$ must therefore be ruled out.

The hypergeometric functions on the right hand side of equation
(\ref{hyperidentity}) are power series in $y^2$ and since $y \propto
{w^n}$ they are power series in $w$.  However, due to the expression
$y^{1-2k}$ the right hand side of equation (\ref{hyperidentity}) will
not consist of integer powers of $w$ unless $k=m/(2n)$ for some
integer $m$.  We now consider the possible values of $m$.  The cases
$m=n$ ($k=1/2$), $m=2n$ ($k=1$) and $m=(2p-1)n$ for $p$ integer ($k$
odd half-integer) have already been ruled out. In considering the
other cases, it is helpful to introduce the quantity ${c_3} \equiv
{c_2} \Gamma ( k-{\textstyle {1 \over 2}}) / ({\sqrt \pi} \, \Gamma
(k))$.  The solution for $H(x)$ near $x=0$ is then
\begin{eqnarray}
H&=&-{2cc_1\over 1+x^n}\left(1-x^{n-m}\right) +O(1) +
c_3\left[x^{n-m}+O(x^{3n-m})\right] \nonumber \\
&=&(c_3-2cc1)x^{n-m}+2cc_1x^{2n-m}+O(x^{3n-m})+O(1) .
\end{eqnarray}
For $m<n$ these are all positive powers of $x$. For $n<m<2n$, we have
one negative power, but it is canceled when $c_1={c_3}/(2c)$. This
means that there is only this one gauge choice $c_1$ in which this
perturbation is regular. For $m>2n$ we have a negative power
$x^{2n-m}$ which cannot be canceled, and we must therefore have
$m<2n$. 

We now consider the perturbation of equation (\ref{esb}) which
(after some straightforward but tedious algebra) becomes
\begin{eqnarray}
\nonumber ny{{d a} \over {dy}} = -{{ 1}
\over { {{(1+y)}^2}}} \biggl [ {c_1} k (1-2n) (1-{y^2}){y^{-2k}} +
{c_1} 2 n k (1-k) {{(1+y)}^2} {y^{-2k}} \\ + {c_1} (2n-1) ( y -
{y^{1-2k}}) + (1-2n) (1-{y^2}) (1+y) {{c_2} \over {2 c}} {F'} \biggr ].
\label{deltaaresult}
\end{eqnarray}
Here $F'$ is an abbreviation for $(d/dy) F(k,1/2,1,1-{y^2})$.

Regularity of the perturbation imposes the condition that
the quantity inside square brackets in
equation (\ref{deltaaresult}) vanishes at $y=0$.  Therefore, for each
term we need only consider that part of it that does not automatically
vanish in this limit.  We first consider the case where $m<n$.  Then,
$y^{1-2k}$ vanishes at $y=0$ but $y^{-2k}$ does not.  Expanding the
quantity inside square brackets in equation (\ref{deltaaresult}) to
the appropriate order we find
\begin{equation}
ny{{d a} \over {dy}} = -{{ 1} \over { {{(1+y)}^2}}} \left [ {c_1} k
(1-2nk) {y^{-2k}} + {{c_3}\over {2 c}} (1-2n) (1-2k) {y^{-2k}} + o(y)
\right ].
\label{mltn}
\end{equation}
Here $o(y)$ denotes a quantity that vanishes when $y=0$.  In general,
the constant $c_3$ can be chosen so that the quantity inside square
brackets in equation (\ref{mltn}) is $o(y)$.  However, for $2nk=1$
({\it i.e.} $m=1$) the choice of $c_3$ that does this is ${c_3}=0$ and
thus the perturbation is pure gauge.  We then arrive at the result
that a physical perturbation cannot have $m=1$.

Now consider the case $n<m<2n$.  Then both $y^{-2k}$ and $y^{1-2k}$
are nonvanishing as $y \to 0$.  In addition, we have the restriction
that ${c_3}= 2 c {c_1}$.  Then expanding the quantity inside square
brackets in equation (\ref{deltaaresult}) to the appropriate order we
find
\begin{equation}
ny{{d a} \over {dy}} =
{{  {c_1} (k-1)} \over {{{(1+y)}^2}}} ( 2nk + 1 - 2n) \left [ {y^{-2k}} + 2
{y^{1-2k}} + o(y) \right ]
\label{mgtn}
\end{equation}
Thus, the only way to obtain regularity of the perturbation is to have
$2nk=2n-1$ or in other words $m=2n-1$.  Note that there is a mode with
$m=2n-1$ only when this value is greater than $n$, that is for $n>1$.

Regularity at the origin $y=1$ does not give rise to further
regularity conditions beyond the ones already discussed.

\section{Conclusions}

Perturbation modes depend on $T$ as $e^{kT}$. As a result of imposing
regularity conditions, we have found the discrete spectrum of growing
perturbations
$k=m/(2n)$, where $m$ is an integer with the following
restrictions: $m>1$ and either $m=2n-1$, or $m<n$.
This means that the CSS background
solution with index $n$ has $n-1$ unstable ($k>0$) modes. In
particular, the solution with $n=1$ has no unstable modes. The $n=2$
solution has one unstable mode with $m=3$. The solutions with $n>2$
all have one unstable mode with $m=2n-1$ and $n-2$ additional unstable
modes with $1<m<n$. 

We now compare our results with the numerical results of
Ref. \cite{mattandfrans}.  The critical solution should be the one
with one unstable mode.  Thus our analysis indicates that the critical
solution is the one with $n=2$ and therefore that $k=3/4$.  In
contrast, the analysis of curvature scaling in \cite{mattandfrans}
indicates that $\gamma = 1.2 \pm 0.05$ which corresponds to $k=0.83
\pm 0.04$.  Furthermore, direct comparison between numerical and
analytic critical solutions indicates a much better fit with $n=4$
than with $n=2$.  The $n=4$ solution has an unstable mode with
$k=7/8=0.875$ which is in agreement with the $k$ of
Ref. \cite{mattandfrans}.  However, our analysis indicates that it
also has modes with $k=1/4$ and $k=3/8$ and therefore this solution
cannot be the critical solution.  Similarly, the $n=3$ solution has an
unstable mode with $k=5/6$ which corresponds to a $\gamma$ of exactly
$1.2$ but it also has a unstable mode with $k=1/3$ and therefore it
seems that it cannot be the critical solution.

Alternatively, it may be that some additional condition needs to be imposed 
on the perturbations due 
to the fact that the past light cone of the singularity 
is an apparent horizon.  Recall that the critical solution is only equal
to our background solution within the past light cone of the singularity
and must be something else outside the past light cone in order to avoid
trapped surfaces.  Thus this critical collapse situation differs from the 
usual one, and it may be that smoothness of the perturbation is not a 
sufficiently strong condition to impose.
Such an additional condition might eliminate some of the 
modes that we have found.
Some light could be shed on this issue by further 
numerical treatment of the
original critical collapse problem.  In particular, such 
a treatment could find the
growing mode and compare it to the modes found in our perturbative treatment.

\acknowledgments

We would like to thank Sean Hayward and Frans Pretorius for helpful 
discussions. This work was partially supported by NSF grant
PHY-9988790 to Oakland University.

\end{document}